\documentclass[10pt]{article}
\usepackage{latexsym}
\usepackage{amsfonts,amsmath,amssymb}
\usepackage{color}
\usepackage{epsfig}
\psfigdriver{dvips}
\usepackage{graphics}
\usepackage{url}

\newtheorem{theoreme}{Theorem}
\newtheorem{proposition}{Proposition}
\newtheorem{lemme}{Lemma}

\newcommand{\qed}{\ensuremath{\hfill\blacksquare}}

\newcommand{\R}{\ensuremath{\mathbb{R}}}
\newcommand{\C}{\ensuremath{\mathbb{C}}}

\newcommand{\E}{\ensuremath{\mathbf{E}}}

\newcommand{\<}{\ensuremath{\langle}}
\renewcommand{\>}{\ensuremath{\rangle}}
\newcommand{\tr}{\mathrm{tr} }
\newcommand{\vol}{\mathrm{vol}}
\newcommand{\Hi}{\ensuremath{\mathcal{H}}}
\newcommand{\co}{\mathrm{conv} }
\newcommand{\Lo}{\mathrm{\textnormal{L\"ow}}}
\newcommand{\John}{\mathrm{\textnormal{John}}}

\newcommand{\Id}{\mathrm{Id}}
\newcommand{\saop}{\mathcal{B}_{sa}(\Hi)}
\newcommand{\vrad}{\mathrm{vrad}}
\newcommand{\e}{\varepsilon}
\renewcommand{\leq}{\leqslant}
\renewcommand{\geq}{\geqslant}
\renewcommand{\circ}{\mathrm{o}}
\renewcommand{\theta}{\vartheta}

\title{Tensor products of convex sets and the volume of separable
states on $N$ qudits }
\author{Guillaume Aubrun  and Stanis\l aw J. Szarek
}
\begin{document}

\date{}
\maketitle

\begin{abstract}
This note deals with estimating the volume of the set of separable
mixed quantum states when the dimension of the state space grows
to infinity. This has been studied recently for qubits; here we
consider larger particles and conclude that, in all cases, the 
proportion of the states that are separable is super-exponentially 
small in the dimension of the set. 
We also show that the partial transpose
criterion becomes imprecise when the dimension increases, and that 
the lower bound $6^{-N/2}$ on the (Hilbert-Schmidt) inradius of the set of 
separable states on $N$ qubits obtained recently by Gurvits and Barnum 
is essentially optimal.  We employ standard tools of classical convexity, 
high-dimensional probability and geometry of Banach spaces.
One relatively non-standard point is a formal introduction of
the concept of projective tensor products of convex bodies, and an
initial study of this concept.
\end{abstract}

\section{Introduction and summary of results}

An important problem in quantum information theory is to estimate
quantitatively parameters related to {\em entanglement}. This
phenomenon is thought to be at the heart of quantum information
processing while, on the other hand, its experimental creation and
handling are still challenging. The question of determining the volume of
the set of separable (i.e., unentangled) 
states has been asked for example in
\cite{zhsl, hhh},  where we refer the reader for background and
motivation. In this paper  we obtain estimates which are meaningful when
the dimension of the state space is ``not too small," and which are
asymptotically tight as that dimension tends to infinity. The effective
radius in the sense of volume of (that is, the radius of the Euclidean
ball of the same volume as; also referred to as {\em volume radius})
the set of separable states for qubits
has been determined (precisely on the scale of powers of the dimension) in
\cite{szarek}. Here we shall deal with more general particles, while
using the same general approach via the methods of asymptotic convex
geometry. The conclusion is that when the dimension of the state space is
large, then all but extremely few (as measured by the standard volume)
states are entangled; see Theorems 1 and 2 below for precise statements. 
As further application of our techniques we show the 
optimality (up to factors of lower order) of the lower bound on the radius
of a Hilbert-Schmidt ball contained in the set of separable states on $N$
qubits that was obtained recently by Gurvits and Barnum \cite{gb3}.
Still another consequence is that, for large dimensions, the set of
states with positive partial transpose \cite{peres, hhh} is, in the sense
of the volume radius, much closer to the set of all states than to the
set of separable states; in other words, the Peres partial
transpose criterion becomes less and less precise as the dimension
increases.

\medskip
We now recall the mathematical framework and introduce some notation.
The state space is a complex Hilbert space $\Hi = \C^{D_1} \otimes \dots
\otimes
\C^{D_N}$. We write $d = D_1 \dots D_N$ for the dimension of
$\Hi$. This Hilbert space allows to describe quantum interactions between
$N$ particles; $D_j=2$ corresponds to qubits and $D_j=3$ to
qutrits. We write $\mathcal{B}(\Hi)$ for the space (or $C^*$-algebra) of linear
maps from $\Hi$ into itself, and $\saop$ for the (real linear) subspace of
self-adjoint operators. The space $\mathcal{B}(\Hi)$ is endowed with the
Hilbert-Schmidt, or Frobenius, scalar product defined by
$\langle A,B \rangle_{HS} := \tr A^\dagger B$  ( $= \sum_{i,j=1}^d
\overline{A_{ij}}B_{ij}$ if $A$ and $B$ are represented as matrices).
A state on $\mathcal{B}(\Hi)$ (which we will abbreviate to ``a state on $\Hi$")
can be represented as a positive (semi-definite) trace one element of $\saop$. A
state is said to be {\em separable } if it can be written as a convex combination
of tensor products of $N$ states, otherwise it is called {\em entangled}. That
is, the set of states (also called in this context density operators or density
matrices) is
$$ \mathcal{D} = \mathcal{D}(\Hi) := \{ \rho \in \saop , \rho
\geq 0, \tr \rho
=1\} ,$$
and the set of separable states is
$$ \mathcal{S} = \mathcal{S}(\Hi) := \co \{ \rho_1 \otimes \dots \otimes
\rho_N, \rho_j \in \mathcal{D}(\C^{D_j}) \} .$$  
The notation $\mathcal{S}(\Hi)$ is in principle ambiguous:
separability of a state  on $\mathcal{B}(\Hi)$ is not an intrinsic property of
the Hilbert space $\Hi$  or of the algebra $\mathcal{B}(\Hi)$; it
depends on the particular decomposition of $\Hi$ as a tensor
product of (smaller) Hilbert spaces. However, this will not be an issue here
since our study focuses on fixed decompositions.

$\mathcal{D}$ and $\mathcal{S}$ are convex
subsets of $\saop$ of (real) dimension $n:=d^2-1$. We write
$\mathcal{T}_1$ for the affine subspace generated by $\mathcal{D}$
(or $\mathcal{S}$); it is the hyperplane of trace one matrices. The
space $\saop$ inherits a (real) Euclidean structure from the
scalar product $\< \cdot,\cdot \>_{HS}$. The corresponding unit ball
will be denoted by $B_{HS}$ (or $B_{HS}^d$). We denote by
$\vol$ the corresponding $d^2$-dimensional Lebesgue volume on $\saop$; we
will also write
$\vol$ for the $n$-dimensional volume induced on $\mathcal{T}_1$.

We shall concentrate on the special case when all the subsystems are
identical, i.e., $D_1=\dots=D_N=D$. Our techniques also apply to
general state spaces, but formulae which fit all the situations
would be more cumbersome than for this ``homogeneous" case. If
$\Hi = (\C^D)^{\otimes N}$, there are two ``regular" ways to make its
dimension tend to infinity: either $N \geq 2$ is fixed
and $D$ tends to infinity, or  $D \geq 2$ is fixed and $N$ goes to
infinity (as in \cite{szarek} for $D=2$). Surprisingly, it turns out that in the
latter case the natural Euclidean structure is not the one given by the
Hilbert-Schmidt scalar product, but rather by a mixture of that product
with a Killing-type form. Our goal is to prove the following theorems
(recall that we work with
$\Hi=(\C^D)^{\otimes N}$ and that $d =
\dim \Hi = D^N$, $N=\log_D{d}$ and
$n=\dim \mathcal{D}=d^2-1=D^{2N}-1$).

\begin{theoreme} [Small number of large subsystems]
There exist universal constants \, $C, c>0$ such that, for all $D, N \geq 2$,
$$
\frac{c^N}{d^{1/2-1/2N}} \leq \left(
\frac{\vol(\mathcal{S})}{\vol (\mathcal{D})} \right)^{1/n} \leq
\frac{C (N\ln N)^{1/2}}{d^{1/2-1/2N}}
$$
\end{theoreme}
We point out that considering the $n$th root of the ratio of volumes is
natural from the geometric point of view: if the two sets had been
Euclidean balls, we would have obtained the ratio of their radii, and so
we are comparing the volume radii of $\mathcal{S}$ and $\mathcal{D}$ 
(see section \ref{urysohn} for additional geometric background). 

\begin{theoreme} [Large number of small subsystems]
There exist universal constants \, $ C', c'>0$ such that, for all $D, N
\geq 2$,
$$ \frac{c'}{d^{1/2+\alpha_D}} \leq \left( \frac{\vol(\mathcal{S})}{\vol
(\mathcal{D})} \right)^{1/n} \leq 
\frac{C'(DN\ln N)^{1/2}}{d^{1/2+\alpha_D}}
$$
where 
$$\alpha_D := \frac{1}{2} \log_D
(1+\frac{1}{D}) -\frac{1}{2D^2}\log_D(D+1)$$
\end{theoreme}
For illustration, we list approximate values of $\alpha_D$ for small $D$'s:
$\alpha_2 \approx 0.094, \alpha_3 \approx 0.061, \alpha_4
\approx 0.044$. Note that $\alpha_D$ is a positive decreasing
function of $D$, asymptotically equivalent to $1/(2D \ln D)$. The quantity 
$d^{1/2+\alpha_D}$ may be alternatively expressed as 
%$\left((D+1)^{1-1/D^2}\right)^{N/2}$.
$((D+1)^{1-1/D^2})^{N/2}$.

\smallskip
Theorem 1 asserts that, as $D\rightarrow \infty$, the {\em ratio of the
volume radii} of $\mathcal{S}$ and $\mathcal{D}$ is of order
$d^{-(1/2-1/2N)}=D^{-(N-1)/2}$  (up to a multiplicative constant depending only
on
$N$).  Theorem 2 is slightly less definite: while, for fixed $D$, the ratio in
question is determined precisely on the scale of the powers of $d$, there
are some (possibly parasitic) logarithmic factors. [We recall again that
$N=\log_D{d} = \ln{d}/\ln{D}$.] We point out that, apart from the value of the
numerical constant,  the upper estimate in Theorem 2 is always sharper than
that in  Theorem 1. However, the gain
$\alpha_D$ in the exponent is negligible  if $N$ is fixed and $D
\rightarrow \infty$. 

It is possible to give reasonable estimates on the constants
appearing in both theorems. One can, for example, take
 $c=1/\sqrt{6}$, $c'=0.3$ and $C=C'=4.4$. It is also possible
to establish a tighter asymptotic behavior 
of these constants (relevant if one is only
interested in large values of $d$ and particularly $N$): one can have 
$c'$ tending to $e^{3/4}/\sqrt{2\pi} \approx 0.845$ as $D$ or $N$ tend to
infinity, and
$C$ and $C'$ tending to $\sqrt{2} \, e^{1/4} \approx 1.816$ as $N$ tends
to infinity.  Finally, sharper estimates for specific dimensions  can be
obtained by using known results on, among others, efficient  spherical
codes; see Appendix G of \cite{szarek}  for an example of such
calculation.

The case $D=2$ of Theorem 2 was treated in detail in \cite{szarek}. In this
note we present additional ingredients required to deal with the
general case. We point out that the (relatively) easy part is the estimation of
$\vol(\mathcal{D})$, which of course does not depend on the tensor
structure of $\mathcal{H}$. It has been shown in \cite{zs} that
\begin{equation}
\vol(\mathcal{D}) = \sqrt{d}\,(2\pi)^{d(d-1)/2}
\frac{\Gamma(1)\dots\Gamma(d)}{\Gamma(d^2)}
\label{explicit}
\end{equation}
A direct
calculation then yields that the volume radius of $\mathcal{D}$
behaves as $e^{-1/4}d^{-1/2}(1+O(1/d))$ as $d\rightarrow \infty$. It is also
possible (and easy) to estimate the volume radius of $\mathcal{D}$  
using techniques with the same flavor as the ones presented
here.  For example, it was shown in \cite{szarek} that for any value of
$d$ that radius is contained between $d^{-1/2}/2$ and $2d^{-1/2}$ (see section
\ref{symmetrization} below). Finally, the fact that the volume radius 
in question is of order $d^{-1/2}$  as $d \rightarrow 
\infty$ follows from an early paper \cite{st}.

\medskip
The ratio of Euclidean volumes is not the only geometric parameter which
is of interest here. 
An arguably more relevant parameter would be its analogue for the
so-called {\em Bures volume}, which may be a more appropriate
measure of size in the present context (see \cite{BC}).
Another quantity with physical significance is the
largest number $\e=\e(\Hi)$ below which the mixture $(1-\e)\Id/d
+ \e \rho$ is separable for any state $\rho$. Of course it is
possible to derive from volume estimates an upper bound on this
$\e(\Hi)$, but this bound is weaker that what is already known
(\cite{qudit, gb, gb3}). 
Still, the knowledge of the volume radii yields additional information 
also in this context: for $\e \gg \left(\vol (\mathcal{S})/\vol
(\mathcal{D})\right)^{1/\dim \mathcal{D}}$, the mixture $(1-\e)\Id/d + \e \rho$
is entangled for ``most of" 
$\rho \in \mathcal{D}$, and not just for very special states as those 
constructed in \cite{qudit, gb, gb3}.
 Another quantity that has been
studied in \cite{gb,gb3,gb1} is the inradius of
$\mathcal{S}$, which is the radius of the inscribed Euclidean ball
in $\mathcal{S}$ (here we assume that the balls are centered at
$\Id/d$, which is the natural center of both $\mathcal{S}$ and
$\mathcal{D}$). It seems that, prior to results on qubits in
\cite{szarek}, the only upper bound which was known on the inradius of
$\mathcal{S}$ was the inradius of $\mathcal{D}$, or $1/\sqrt{d(d-1)}$. 
[A remarkable fact discovered in \cite{gb1} is that the two inradii coincide
when $N=2$.] 
Theorem 2  improves on this for large $N$ giving, up to logarithmic factors, 
the upper bound $1/d^{1+\alpha_D}$. Moreover, the argument shows that
the same is true for the symmetrized body 
$\Sigma := \co (\mathcal{S} \cup -\mathcal{S})$.
[Cf. section \ref{symmetrization} and note that the distance from the
hyperplane containing 
$\mathcal{S}$ to the origin is $1/\sqrt{d} > 1/\sqrt{d(d-1)}$, so the
inradius of
$\Sigma$ is at least as large as that of $\mathcal{S}$.]
The techniques developed in
\cite{szarek}  and in the present paper allow to tighten this bound
substantially. We present the following sample result.

\begin{theoreme} [Tight upper bounds on the inradii of $\mathcal{S}$ and
$\Sigma$] There exists an absolute constant $C_0>0$ such that, for any $N$,
the set $\Sigma((\C^2)^{\otimes N})$, and hence also
$\mathcal{S}((\C^2)^{\otimes N})$, does not contain any Hilbert-Schmidt
ball of radius $C_0\sqrt{N \ln{N}} \; 6^{-N/2}$.
\end{theoreme}
For comparison, let us cite the lower bound of $6^{-N/2}$ on the inradius 
of the set $\Sigma$ noted in \cite{szarek}, Appendix H, and a similar lower
bound on the {\em a priori } smaller inradius of $\mathcal{S}$ shown
subsequently in \cite{gb3}. This means that, at least for $N$ qubits, the
inradii of $\mathcal{S}$ and of $\Sigma$ have been determined up to a 
multiplicative factor which is logarithmic in the dimension of the state
space. The argument we present can be generalized to other values of $D$,
leading to an upper bound for the inradii of $\Sigma((\C^D)^{ \otimes N})$
and $\mathcal{S}((\C^D)^{\otimes N})$ which is $(D(D+1))^{-N/2}$ (again, 
up to a logarithmic factor). For the inradius of the set $\Sigma$ this 
bound is essentially sharp. However, for $D>2$, the value
$(D(D+1))^{-N/2}$ does not coincide with the lower bound for the inradius of
$\mathcal{S}((\C^2)^{\otimes N})$ that was found in
\cite{gb3} and which is of the form  $O((D(2D-1))^{-N/2})$.
As was noticed in \cite{gb3}, a better upper bound for $D>2$ is derived 
from our upper bound for $D=2$ and is (up to a logarithmic
factor) of the form $(1.5 \times D^2)^{-N/2}$. However, this still
leaves a gap between the upper and lower estimates which is
exponential in $N$ (as $N \rightarrow \infty$).
A similar discussion of the bilateral case ($N=2, D\rightarrow \infty$) 
can be found in the Remark at the end of section 3.

\medskip

Theorem 1 above means that {\em typical} (as measured by
the standard volume) high-dimensional bipartite states are entangled. This
statement is of global nature, and is therefore not helpful  when one
faces a given state and has to decide about its separability. In turn,
{\em lower } bounds for the inradii of sets $\Sigma$,  or for the
quantity $\e(\Hi)$ mentioned above provide a {\em sufficient } condition
for separability which, while very useful in some contexts, is
conclusive  for only a very small fraction of separable states. A
very simple and efficient (at least in small dimensions) 
{\em necessary } condition is the so-called {\em Peres partial transpose
criterion} (\cite{peres}). We briefly recall what partial transposition
is, say when the state space is
$\mathcal{H} = \C^D
\otimes
\C^D$.  Fix a basis $(e_1,\dots,e_D)$ of $\C^D$, for example the canonical
basis. Any state $\rho$ on $\mathcal{H}$ can be uniquely expressed as
$$\rho = \sum_{i,j,\alpha,\beta=1}^D \rho_{i\alpha,j\beta}|e_i\otimes
e_\alpha \rangle \langle e_j \otimes e_\beta |$$ We then consider
$T\rho$, the partial transpose of $\rho$ (with respect to the
first subsystem), defined by the formula
$$ (T\rho)_{i\alpha,j\beta} = \rho_{j\alpha, i\beta} $$
One checks that $T\rho$ is also an Hermitian operator of trace one which,
however, is not necessarily  positive. Following
\cite{hhh}, we write $\mathrm{PPT}$ (Positive Partial Transpose)
for the set of states $\rho$ for which $T\rho$ is also positive.
Note that while $T\rho$ depends on the choice of the basis, its
eigenvalues do not; so the set $\mathrm{PPT}$ is
basis-independent. The Peres criterion simply reads: {\em any
separable state has a positive partial transpose}. In other words,
we have the inclusions $\mathcal{S} \subset \mathrm{PPT} \subset
\mathcal{D}$. An natural question (asked for example in \cite{hhh}) is
then:  {\em what is the volume of
$\mathrm{PPT}$?} Is it close to the volume of $\mathcal{S}$, or
rather to the volume of $\mathcal{D}$? The following theorem
answers the latter question.

\begin{theoreme} [Asymptotic roughness of PPT criterion]
There exists an absolute constant $c_0>0$ such that for any
bipartite system $(\Hi=\C^D \otimes \C^D)$
$$c_0 \leq \left( \frac{\vol(\mathrm{PPT})}{\vol(\mathcal{D})}
\right)^{1/n}
$$
\end{theoreme}

An immediate corollary to Theorems 1 and 4 is that, for large $D$, the
volume of PPT states is much bigger than the volume of separable states.
This means that a typical high-dimensional PPT state is entangled, and
thus that Peres criterion becomes less and less precise as the dimension
increases. Theorem 4 is just a sample result; similar analysis may be
performed for other bipartite and multipartite systems. On the other hand,
it is quite conceivable that the quantity from Theorem 4 admits a
nontrivial (i.e., $<1$, independent of $D$) upper bound, even in the case
of bipartite systems. [This would imply that vast majority of
entangled states are detected by the PPT criterion.]

\section{Preliminaries about convex bodies}

In this section we list some known (or elementary) facts from convexity
theory that will be needed in the sequel. A {\em convex body} is a convex
compact subset of $\R^n$ (or of $\C^m$ identified with $\R^{2m}$, 
or of an {\em affine} subspace of
$\R^n$) with nonempty interior. We will say that a convex body $K$ is {\em
symmetric} if it is symmetric with respect to the origin (i.e., $K=-K$). To a
convex body $K \subset \R^n$  containing the origin in its interior one can
canonically associate $\|\cdot\|_K$, the {\em gauge of} $K$, by setting, for
$x \in \R^n$,  $\|x\|_K := \inf \{\lambda > 0 \ : \  x\in \lambda K \}$. 
If $K$ is symmetric, $\|\cdot\|_K$ is the norm for which $K$ is the unit
ball. If $K$
and $L$ are two convex bodies in $\R^n$, each of which contains the origin
in its interior, we define the {\em geometric distance} between
$K$ and $L$ as the product $\alpha\beta$, where $\alpha$ and
$\beta$ are the smallest numbers for which $\beta^{-1} L \subset K \subset
\alpha L$ . The concept generalizes to bodies with a common interior 
point contained in an affine subspace of $\R^n$; it
quantifies how the shapes of $K$ and $L$ differ.

\subsection{The Urysohn inequality} \label{urysohn}

Let $K \subset \R^n$ be a convex body and $u \in S^{n-1}$ be a
unit vector. The {\em width of $K$} in the direction $u$ is
$$ w(K,u) := \max_{x \in K} \langle u,x \rangle $$
Geometrically, $w(K,u)$ is (up to the sign) the distance from the origin
to the hyperplane tangent to $K$ in the direction $u$. The {\em mean
width of
$K$} is simply the mean of the widths in all directions: $$ w(K) :=
\int_{S^{n-1}} w(K,u) d\sigma(u) $$ where $\sigma$ is the Lebesgue
measure on the sphere, normalized so that $\sigma(S^{n-1}) =1$.
The classical {\em Urysohn inequality}, which is of isoperimetric flavor,
states that among all convex bodies of given volume, the Euclidean ball
has the minimal mean width. It can be written as follows (see, e.g.,
Appendix A of \cite{szarek} for a proof)
$$ \left(\frac{\vol (K)}{\vol(B^2_n)}\right)^{\frac{1}{n}} \leq w(K) ,$$
where $B_2^n$ is the $n$-dimensional Euclidean unit ball. The quantity on
the left-hand side equals  the radius of the Euclidean ball which has the
same volume as $K$ and, as we mentioned earlier, is sometimes called the
{\em volume radius} of $K$; we will denote it $\vrad (K)$.
We will use the Urysohn inequality to estimate, among others, 
the volume radius of
$\Sigma$, which is the quantity we are interested in. It is often more 
convenient to rewrite the inequality above by expressing the mean width 
via a Gaussian integral (rather than an integral over the
sphere):
$$ \vrad(K) \leq \frac{1}{\gamma_n} \E \max_{x \in K} \langle x,G
\rangle ,$$
where $G$ is a $\mathcal{N}(0,\Id_n)$ Gaussian vector, $\E$ is the
expectation with respect to $G$ and $\gamma_n := \sqrt{2}
\Gamma(\frac{n+1}{2})/\Gamma(\frac{n}{2})$ is a constant which is close to
$\sqrt{n}$.  The identity  
$ w(K) = \gamma_n^{-1} \E \max_{x \in K} \langle x,G\rangle$ 
is checked by passing to polar coordinates. We will occasionally call   
$\E \max_{x \in K} \langle x,G\rangle$ the {\em Gaussian mean width } 
of $K$ and denote it $w_G(K)$.

A special situation is when the body $K$ is a polytope with $v$
vertices on the unit sphere. In this case it is well-known 
(see  Proposition 1.1.3. in \cite{talagrand} for a nice proof) that
the expectation $w_G(K)$ is smaller than $\sqrt{2\ln v}$. A somewhat more
complicated argument based on another proof (\cite{fernique}, Lemme 0.6.2) 
and supplemented by numerics shows that the same estimate works for a {\em
symmetric } polytope with $2v$ vertices on the unit sphere (or just in the
unit ball) provided $v>1$. We eventually obtain
 that, for $v>1$ points $(x_i)_{i=1}^v$ in the unit ball,
\begin{equation}
\vrad (\co \{\pm x_i\}_{i=1}^v) \leq w(\co \{\pm x_i\}_{i=1}^v) 
<  \frac{1}{\gamma_n}
\sqrt{2\ln v} .
\label{volpol}\end{equation}
Note that since $\gamma_n = w_G(B_2^n) \sim \sqrt{n}$, 
the quantity on the right
will be small as long as $\ln{v} \ll n$.

\subsection{Symmetrization and the Rogers-Shephard inequality}
\label{symmetrization}

Let $H$ be an affine hyperplane in $\R^{n+1}$, not containing the origin, 
and let $W$ be a convex body in $H$. Consider the symmetrization $\Omega$ 
of $W$, defined by $\Omega := \co (W \cup -W)$.  This procedure is
quite natural in the framework of convex bodies, since much more
information is available about centrally symmetric convex bodies as
opposed to general ones. In particular,  $\Omega$ is the unit ball
with respect to the norm  $\|.\|_\Omega$.
If $h>0$ is the distance between $H$ and
the origin, then we have the following inequalities
\begin{equation}
 2h \, \vol(W) \leq \vol(\Omega) \leq 2h \, \frac{2^n}{n+1}\vol(W) .
\label{rstube}
\end{equation}
The left hand inequality is an immediate consequence of the
{\em Brunn-Minkowski inequality} (\cite{gardner}), and is an equality if
the body $W$ is centrally symmetric. The right hand inequality is the
{\em Rogers-Shephard inequality} (\cite{rs}) and equality is achieved
when $W$ is a simplex. The factor $\frac{2^n}{n+1}$ may appear large,
but we point out that its $n$th root, which is more relevant here, 
is smaller than 2.

We now consider symmetric versions of $\mathcal{D}$ and
$\mathcal{S}$, namely $\Delta := \co (\mathcal{D} \cup
-\mathcal{D})$ and $\Sigma := \co (\mathcal{S} \cup
-\mathcal{S})$ (the latter was defined already in the Introduction); then
$n=d^2-1$. The set
$\Delta$ is just the Hermitian part of the unit ball of the
usual Schatten class of trace class operators. In other words 
$\Delta = \{ A \in \saop , \|A\|_1 \leq 1\}$, where $\|A\|_1$ is the {\em
trace class norm}, i.e., the sum of singular values of $A$ or, in our
Hermitian setting, of the absolute values of the eigenvalues of $A$. [See
section
\ref{tensor} for the functional-analytic interpretation of $\Sigma$.] The
inequalities (\ref{rstube}) imply now 
\begin{equation} \frac{1}{2} \left(\frac{\vol
(\Sigma)}{\vol(\Delta)}\right)^{1/n} \leq \left(\frac{\vol
(\mathcal{S})}{\vol(\mathcal{D})}\right)^{1/n} \leq 2 \left(\frac{\vol
(\Sigma)}{\vol(\Delta)}\right)^{1/n} \label{r-s}
\end{equation}
Therefore, the loss in precision when passing from
the ratio of the volume radii of $\mathcal{S}$ and $\mathcal{D}$ to the
ratio of  the volume radii of $\Sigma$ and $\Delta$ is at most a factor of
2.  To be precise, the exponent obtained for the
symmetrized bodies is $1/n$ whereas it should be $1/d^2$ (which is
the reciprocal of the dimension of $\Delta$ and $\Sigma$), but this is
not at all an issue. First, we can {\it a posteriori} change the
exponent, this will at most slightly affect the constants. Second,
the extra $n+1$ in (\ref{rstube})  ensures that no modification of
the constants is actually needed (see \cite{szarek}, Appendix C, 
for a precise general statement in this direction).

For future reference we recall here the estimates from
\cite{szarek}  for the volume radius of $\mathcal{D}$ (cited in the
Introduction), for the volume radius of $\Delta$, and for the mean
widths of these sets. 
\begin{equation}
\label{voldd}
\frac{1}{\sqrt{d}} \ \leq \ \vrad(\Delta) =
\left(\frac{\vol(\Delta)}{\vol(B_2^{d^2})}\right)^{1/d^2} \leq \ w(\Delta) \ 
\leq \frac{2}{\sqrt{d}}
\end{equation}
\begin{equation}
\label{vold}
\frac{1}{2\sqrt{d}} \ \leq \ \vrad(\mathcal{D}) \leq \ w(\mathcal{D}) \ 
\leq \frac{2}{\sqrt{d}}
\end{equation}
While, as we mentioned in the Introduction, both a closed expression for 
$\vrad(\mathcal{D})$ and its precise asymptotic behavior (as the dimension
increases) are known, and while presumably similar information about 
$\vrad(\Delta)$ may be obtained via the methods of \cite{zs}, the compact and
transparent inequalities from (\ref{voldd})  and (\ref{vold}) are
sufficient for our asymptotic results. Moreover, having upper bounds for the
{\em larger} parameter $w(\cdot)$ will provide us with additional flexibility.

\subsection{Ellipsoids associated to convex bodies}\label{ellipsoids}

An ellipsoid is an affine image of the unit ball $B_2^n$. There is
a one-to-one correspondence between symmetric ellipsoids and
scalar products (or Euclidean structures). Indeed, if
$\mathcal{E}=TB_2^n$, with $T$ a linear map, then $\mathcal{E}$ is
the unit ball associated to the scalar product
$\<x,y\>_{\mathcal{E}} := \<T^{-1}x,T^{-1}y\>$. If $K$ is a convex
body in $\R^n$, its {\em polar body}, denoted $K^\circ$, is defined by
$K^\circ=\{ x \in \R^n, \forall y \in K, \<x,y\> \leq 1 \}$.  
If $0$ belongs to the interior of $K$, $K^\circ$ is also a convex body 
(if  not, it fails to be bounded). 
We point out that $\|x\|_{K^\circ} = \max_{y \in K} \<x,y\>$ and that the
restriction of the gauge $\| \cdot \|_{K^\circ}$ to the sphere coincides 
with $w(K,\cdot)$, the width function of $K$ defined in section \ref{urysohn}.
We further note that the polar of a symmetric ellipsoid is again a symmetric
ellipsoid. The following definition is useful for our purposes.

\medskip

{\bf Definition: } If $K$ is a convex body in $\R^n$, a {\it John
resolution of identity} associated to $K$ is a finite family
$(x_i,c_i)_{i \in I}$, where $x_i$ belong to $\partial K \cap
S^{n-1}$ (if $K \subset B_2^n$ or $B_2^n \subset K$, these are
contact points of $K$ with the sphere) and $c_i$ are positive
numbers, such that
\begin{enumerate}
\item $\sum c_i x_i =0$
\item $\forall y \in \R^n, y=\sum c_i \<x_i,y\>x_i$
\end{enumerate}
Condition 2 can be rephrased using Dirac notation as $\sum c_i |
x_i \> \< x_i | = \Id_n $. Taking trace both sides, we see that
necessarily $\sum c_i=n$. We recall the following result due to John

\begin{proposition}[John's theorem]
\label{john} Let $K$ be a convex body in $\R^n$. Then there exists
a unique ellipsoid of minimal volume containing $K$, called
L\"owner ellipsoid, which we denote $\Lo(K)$. Similarly, there
exists a unique ellipsoid of maximal volume contained in $K$, called John
ellipsoid, which we denote $\textnormal{John}(K)$. Moreover, the
equality $\Lo(K)=B_2^n$ holds if and only if $K \subset B_2^n$ and
there exists a John resolution of identity associated to $K$.
Similarly, the equality $\textnormal{John}(K)=B_2^n$ holds if and
only if $B_2^n \subset K$ and there exists a John resolution of
identity associated to $K$.
\end{proposition}

For a proof of this theorem, see \cite{ball}. These ellipsoids are
affine invariants: for any affine map $T$, we have
$\John(TK)=T\John(K)$ and $\Lo(TK)=T\Lo(K)$. In the symmetric case
(the one we mostly deal with in the sequel), John and L\"owner ellipsoids
are also symmetric and dual to each other with respect to
polarity. More precisely, if $K$ is symmetric, we have
$\John(K^\circ) = \Lo (K)^\circ$ and $\Lo(K^\circ) =
\John(K)^\circ$; this is a consequence of the fact that for any
symmetric ellipsoid $\mathcal{E}$, we have  $\vol
(\mathcal{E}^\circ) = \vol (B_2^n)^2 / \vol(\mathcal{E})$. It is
therefore immediate to pass from the statement about John
ellipsoid to the statement about L\"owner ellipsoid (actually the
theorem is usually stated for the John ellipsoid only).

\subsection{Tensor products of convex bodies} \label{tensor}

If $K$ and $K'$ are convex bodies, respectively in
$\R^n$ and $\R^{n'}$, their {\em projective tensor product} is the
convex body $K \hat{\otimes} {K'}$ in $\R^n \otimes \R^{n'} \sim \R^{nn'}
$ defined as follows
$$ K \hat{\otimes} K' = \co \{x \otimes x', x \in K, x' \in K'\} .$$
This terminology is motivated by the fact that when 
$K$ and $K'$ are unit balls with respect to some norms, the set 
$ K \hat{\otimes} K'$ is the unit ball of the corresponding 
projective tensor product norm on $\R^n \otimes \R^{n'}$. The relevance
of the notion to our discussion is obvious: the set of separable states
is the projective tensor product of sets of states on factor spaces; more
precisely, if $\Hi = \Hi_1 \otimes \Hi_2$, then 
$ \mathcal{S}(\Hi) = \mathcal{D}(\Hi_1) \hat{\otimes} \mathcal{D}(\Hi_2)$.
It is easy to see that the operation $\hat{\otimes}$ commutes
with symmetrizations: if $\tilde{K} =\co (K \cup -K)$ and $\tilde{K}'
=\co (K'\cup -K')$, then 
$\co (K \hat{\otimes} K',-K \hat{\otimes} K')=\tilde{K} \hat{\otimes}
\tilde{K}'$. It follows that, in our notation,   
$ \Sigma(\Hi) = \Delta(\Hi_1) \hat{\otimes} \Delta(\Hi_2)$. Since, as was
pointed out in section \ref{symmetrization}, $\Delta$ is the
unit ball in the trace class norm, $ \Sigma(\Hi)$ is the unit ball of the 
projective tensor product of the spaces ${\mathcal{B}_{sa}(\Hi_k)}$,
$k=1,2$, each endowed with the trace class norm.

\smallskip 
The definition of $ K \hat{\otimes} K'$ -- similarly as other
definitions, comments and lemmas of this section -- immediately generalizes to
tensor products of any finite number of factors. However, for the sake of
transparency, we shall concentrate in this section on the case of two bodies.
We also point out that while the definition appears to be well-adapted to
0-symmetric sets, cones and linear maps as morphisms (this is used in the
proof of Lemma \ref{lemtens} below and, later, in the first paragraph of
section 5), the projective tensor product {\em is not } invariant under 
affine maps. We refer to \cite{svetlichny}, Chapter 2, for a
discussion of related categorical issues and to \cite{df} for exhaustive
treatment of tensor products of normed spaces.

\smallskip 
If $\mathcal{E}=SB_2^n$ and $\mathcal{E}'=S'B_2^{n'}$ are two
ellipsoids, respectively in $\R^n$ and $\R^{n'}$, their Hilbertian
tensor product is the ellipsoid $\mathcal{E} \otimes_2 \mathcal{E}'
:= (S \otimes S')B_2^{nn'}$. This definition does not depend on the
choice of $S$ and $S'$. It turns out that L\"owner ellipsoids
behave well with respect to projective tensor product, as the
following lemma shows (note that the analogous statement {\em does not }
hold for the John ellipsoid).

\begin{lemme}
\label{lemtens}
Let $K \subset \R^n$ and $K' \subset \R^{n'}$ be two convex
bodies. Then
the L\"owner ellipsoid of their projective tensor product is the
Hilbertian tensor product of the respective L\"owner ellipsoids.

In terms of scalar products, for every $x,y$ in $\R^n$ and $x',y'$ in
$\R^{n'}$, we have
$$ \< x \otimes x',y \otimes y' \>_{\Lo(K\hat{\otimes}K')} =
\<x,y\>_{\Lo(K)} \<x',y'\>_{\Lo(K')} $$
\end{lemme}

{\bf Proof:} First suppose that $\Lo(K)=B_2^n$ and $\Lo(K')=B_2^{n'}$. By
John's theorem, there exist John resolutions of identity for $K$ and $K'$,
respectively $(x_i,c_i)_{1\leq i\leq k}$ and $(x'_j,c'_j)_{1\leq j\leq
k'}$.  We easily check that $K \hat{\otimes} K' \subset B_2^{nn'} = B_2^n
\otimes_2 B_2^{n'}$. Moreover, $(x_i \otimes x'_j,c_ic'_j)_{1\leq i\leq
k,1\leq j\leq k'}$ is also a John resolution of identity for $K
\hat{\otimes}
K'$. It is enough to check this on pure tensors: take $y \in \R^n$ and
$y'
\in \R^{n'}$ and verify the two conditions
\begin{enumerate}
\item $ \sum_{i=1}^k \sum_{j=1}^{k'} c_ic'_j x_i \otimes x_j' =(
\sum_{i=1}^k c_ix_i) \otimes ( \sum_{j=1}^{k'} c'_j x_j' ) =0 $
\item $ \sum_{i=1}^k \sum_{j=1}^{k'} c_ic'_j \<x_i \otimes x'_j,y\otimes
y'\>x_i \otimes x'_j = (\sum_{i=1}^k c_i \<x_i,y\> x_i ) \otimes (
\sum_{j=1}^{k'} c'_j \<x'_j,y'\>x'_j ) = y \otimes y' $
\end{enumerate}
For the general case, let $T$ and $T'$ be linear maps such that
$T\Lo(K)=B_2^n$ and $T'\Lo(K')=B_2^{n'}$. Using the elementary identities 
$\Lo(TK)=T\Lo(K)$ and $(T \otimes T')(K \hat{\otimes} K') = (TK)
\hat{\otimes} (T'K')$, the result follows from the previous particular
case. \qed

\medskip The next ``Chevet-Gordon type" lemma relates the mean widths of
convex sets to that of  their projective tensor product. 
It is most conveniently
stated for  the mean {\em Gaussian } width $w_G(\cdot)$ defined in section
\ref{urysohn}.
\begin{lemme}
\label{tenswidth}
Let $K \subset B_2^n \subset \R^n$ and $K' \subset B_2^{n'} \subset \R^{n'}$
be convex sets, one of which is the convex hull of a subset of the
corresponding unit sphere. Then
$$ w_G(K\hat{\otimes}K') \leq w_G(K) + w_G(K').$$
\end{lemme}
Restating the assertion of the Lemma in terms of the standard mean width gives
$$
w(K\hat{\otimes}K') \leq \frac{\gamma_n}{\gamma_{nn'}}w(K) +
\frac{\gamma_{n'}}{\gamma_{nn'}}w(K') 
\leq \frac{w(K)}{\sqrt{n'}} + \frac{w(K')}{\sqrt{n}} ,
$$
where the second inequality follows from the fact that the sequence
$(\gamma_k/\sqrt{k})$ is increasing (which can be deduced from inequalities
proved in \cite{luke}; the fact that  the coefficients on the right hand
side are {\em approximately} 
$1/\sqrt{n'}$ and $1/\sqrt{n}$ follows from the simpler relationship 
$\gamma_k \sim \sqrt{k}$.) We thus get seemingly strong
upper bounds for mean widths and, {\em a fortiori}, for volume radii of
projective tensor products of convex bodies. However, in spite of
its elegance, Lemma
\ref{tenswidth} will play only rather limited role in our arguments. 
This is because, when iterated,  it yields asymptotically worse dependence 
on the number of factors than the techniques employed in subsequent sections.
Accordingly, its main use will be in considerations involving few factors and
in ameliorating the numerical constants which appear in the statements of the
Theorems. The Lemma can be shown by an argument very similar to that of
section 2.3 of \cite{ds}, we skip the proof; here we just note
that we do not know whether the hypothesis that one of the sets be
spanned by a subset of the sphere is necessary. This requirement is
satisfied for all sets in our applications, and it is not at all
restrictive if the widths of $K, K'$ are estimated via 
cardinalities of their spanning sets.

\smallskip
Finally, let us note that if $K$ and
$K'$ are circled convex bodies in respective complex vector spaces (a convex
body
$K \subset \C^n$ is said to be circled if $K=e^{i\theta}K$ for all
real $\theta$), then $ K \hat{\otimes} K'$ is circled; as was the case
for symmetrizations, the operation $\mathrm{absconv}$ (the absolute
convex hull) commutes with  the projective tensor product.

\section{Proof of Theorem 1: Small number of large subsystems}

The upper bound can be obtained through a standard discretization
argument. Recall that, for $\delta >0$,  a $\delta$-net of a set $K$ is a
subset
$\mathcal{N} \subset K$ such that for each $x \in K$ there exists
$y \in \mathcal{N}$ whose distance to $x$ is smaller than $\delta$. If
$\mathcal{N} \subset \C^D$, 
we write $P(\mathcal{N})$ for the polytope
$\co
\{ \pm |x\rangle \langle x| \}_{x \in \mathcal{N}} \subset
\mathcal{B}_{sa}(\C^D)$. The following elementary lemma shows that if
$\mathcal{N}$ is a net of the unit sphere of $\C^D$ -- which may be
identified with $S^{2D-1}$ as a metric space -- then $P(\mathcal{N})$ is a
good approximation of $\Delta(\C^D)$.

\begin{lemme}
\label{net} Let $\delta < \sqrt{2-\sqrt{2}} \approx 0.765$ and let
$\mathcal{N}$ be a
$\delta$-net of the unit sphere of $\C^D$. Then 
$$ (1-2\delta^2+\delta^4/2)\Delta(\C^D) \subset P(\mathcal{N}) \subset
\Delta(\C^D)$$
\end{lemme}
 
{\bf Proof:} The second inclusion is trivial. Let us check the
first one through the corresponding dual (polar) norms
$$ 
\|A\|_{\Delta(\C^D)^\circ} 
= \max_{x \in \C^D, \|x\|=1}  \< \, A, \pm |x\>\<x| \, \>_{HS}  
= \max_{x \in \C^D, \|x\|=1} \left| \langle x |A | x \rangle \right| 
= \max_{\lambda \in \sigma(A)}|\lambda| = \|A\|_{op} ,
$$
$$ \|A\|_{P(\mathcal{N})^\circ} = 
\max_{y \in \mathcal{N}} \left| \langle
y | A | y \rangle \right| 
$$
 where
$\|\cdot\|_{op}$ is the operator norm and $\sigma(\cdot)$ is the
spectrum (recall that we  consider only Hermitian matrices here). 
We need to show that $ \|A\|_{P(\mathcal{N})^\circ} \ge
(1-2\delta^2+\delta^4/2) \|A\|_{op}$. To this end,
let $A \in \mathcal{B}_{sa}(\C^D)$ be such that $\|A\|_{op}$ and the
largest eigenvalue of $A$  are both equal to 1, and let $x \in \C^D$ be a
norm one vector such that $Ax=x$.  Choose $y_0 \in \mathcal{N}$ verifying
$\|x-y_0\|\leq \delta$.  We claim that 
$\langle y_0 | A | y_0\rangle \ge 1-2\delta^2+\delta^4/2$; the inequality
between the norms follows then by homogeneity. The claim is easily
established by writing $y_0 = y_1+y_2$ with $y_1 = \langle y_0 | x
\rangle x$ and noting that 
$\langle y_1 | A | y_1\rangle \ge \left|\langle y_0 | x \rangle \right|^2
\ge (1-\delta^2/2)^2$, $\left| \langle
y_2 | A | y_2 \rangle \right| \le \|y_2\|^2 \le \delta^2(1-\delta^2/4)$
and $\langle y_1 | A | y_2 \rangle = 0$. \qed

\medskip

Tensoring the conclusion of the preceding lemma and recalling that
$\Delta(\C^D)^{\hat{\otimes}N} = \Sigma(\Hi)$
yields an inclusion $(1-2\delta^2+\delta^4/2)^N
\Sigma(\Hi) \subset P(\mathcal{N})^{\hat{\otimes}N}$. Observe that
$P(\mathcal{N})^{\hat{\otimes}N}$ is another symmetric polytope with at most
$2(\#\mathcal{N})^N$ vertices. It is well-known (\cite{pisier},
Lemma 4.10) that for $\delta \leq 1$ we can
find $\delta$-nets in
$S^{2D-1}$ of cardinality not exceeding  $\leq
(1+2/\delta)^{2D}$. We thus get a bound on the volume of the
polytope using the estimation (\ref{volpol}).

We note for future reference that the above discussion can be
carried out using other Euclidean structures, the only constraint
being that the vertices of the polytope are of norm not exceeding
 1. Thus, if $\mathcal{E}$ is any ellipsoid containing
$\Sigma(\Hi)$, we obtain
\begin{equation}\label{upper}
\left( \frac{\vol(\Sigma(\Hi))}{\vol(\mathcal{E})}\right)^{1/d^2}
\leq \inf_{0<\delta<\sqrt{2-\sqrt{2}}} 
\frac{1}{\gamma_{d^2}(1-2\delta^2+\delta^4/2)^N}
\sqrt{2 \ln \left((1+2/\delta)^{2DN}\right)}
\end{equation}
Using $\mathcal{E}=B_{HS}$ and, say,  $\delta=1/\sqrt{N \ln{2N}}$
leads to an estimate $\vrad(\Sigma(\Hi)) = O((DN \ln{N})^{1/2}/d)$. 
[Indeed, for large
$d$ we have then  $\gamma_{d^2} \sim d$
and, for large $N$,  
$(1-2\delta^2+\delta^4/2)^N \sim 1$ and
$\ln \left((1+2/\delta)^{2DN}\right) \sim DN\ln{N}$.] We obtain the
upper bound announced in Theorem 1 by combining this with the  {\em
lower} bound on 
$\vrad(\Delta(\Hi))$ given by (\ref{voldd})  and with (\ref{r-s}). 
Finally, the more precise statements about the constants indicated in the
introduction can be obtained by using  (\ref{explicit}), Lemma
\ref{tenswidth} and Appendix C of \cite{szarek}.

\smallskip
The lower estimate will be proved by showing that $\Sigma(\Hi)$ contains a
Euclidean ball of appropriately large volume. We start with the following
lemma
\begin{lemme}
\label{inradius} Let $K := (B_2^D)^{\hat{\otimes} m}$ be the
projective tensor product of $m$ copies of the  
$D$-dimensional Euclidean ball (real or complex). Then $K$ contains the
following multiple of the Euclidean ball in $(\C^D)^{\otimes m}$ (resp., 
$(\R^D)^{\otimes m}$)
$$ \frac{1}{D^{(m-1)/2}} B_2^{D^m} \subset K .$$
\end{lemme}

{\bf Proof:} Let $(e_1,\dots,e_D)$ be
the canonical basis of
$\C^D$. We write an element $A$ of $(\C^D)^{\otimes m}$ as a generalized
matrix:
$A=(a_{i_1\dots i_m})$ stands for
$$ A = \sum_{\small \begin{array}{c} i_1, \ldots, i_m=1
\end{array}}^D
a_{i_1 \dots i_m} e_{i_1} \otimes \dots \otimes e_{i_m}
$$

The norm dual to $\|\cdot\|_K$ is 
\begin{equation}
\|A\|_{K^\circ} = \max_{\small \begin{array}{c} x^1,\ldots,
x^m \in B_2^D \end{array}} \left| \sum_{\small \begin{array}{c} i_1,
\ldots ,i_m=1
\end{array}}^D a_{i_1 \dots i_m} x^1_{i_1} \hdots x^m_{i_m} \right|
\label{defnkp}
\end{equation}

Let us write $\|A\|_2$ for the Euclidean norm of $A$, i.e., $\|A\|_2^2 =
\sum |a_{i_1\dots i_m}|^2 $. We want to show that $\forall A \in
(\C^D)^{\otimes m}$ we have $ D^{-(m-1)/2}\|A\|_2 \leq \|A\|_{K^\circ}$. 
By homogeneity, we can assume that $\|A\|_2 =1$. Now choose
$x^1,\hdots,x^{m-1}$ randomly and independently to be uniformly
distributed on the unit sphere of $\C^{D}$. We write $\E$ for the
corresponding expectation. Let $X_k$, $k=1, \dots, D$, be the random variable
obtained by summing only on the $k$th ``hyper-slice,'' that is 
$$ X_k=X_k(x^1,\dots,x^{m-1}) :=  \sum_{\small \begin{array}{c}
i_1, \ldots, i_{m-1}=1 \end{array}}^D a_{i_1\dots i_{m-1}k}x^1_{i_1} \dots
x^{m-1}_{i_{m-1}}$$ We can easily calculate $\E|X_k|^2$ since many
cancellations come from the fact that $\E (x^k_i\overline{x^l_j}) =
\frac{1}{D}\delta_{kl}\delta_{ij}$. We obtain
$$ \E | X_k|^2 = \frac{1}{D^{m-1}}\sum_{\small \begin{array}{c}
i_1,\ldots,i_{m-1}=1 \end{array}}^D \left| a_{i_1 \dots i_{n-1}k}
\right|^2
$$ Therefore, $\E( \sum_{k=1}^D |X_k|^2) = \frac{1}{D^{m-1}}$. This
implies the existence of unit vectors $x^1,\dots,x^{m-1}$ such that,
denoting $Y_k:=X_k(x^1,\dots,x^{m-1})$, we have
$\sum_{k=1}^D |Y_k|^2 \geq \frac{1}{D^{m-1}}$.
Choosing these points in (\ref{defnkp}), we obtain
$$ \|A\|_{K^\circ} \geq \max_{x^m \in B_2^D} \sum_{k=1}^D Y_k x^m_k =
\left(\sum_{k=1}^D |Y_k|^2\right)^{1/2} \geq \frac{1}{D^{(m-1)/2}}$$
\qed

\medskip

{\bf Remark: }It turns out that this lemma is surprisingly sharp. 
As shown in Lemma \ref{vrad} below, even the {\em a priori } larger volume
radius of $K$ is (up to multiplicative factor depending only on $m$) of the
same order
$D^{-(m-1)/2}$. Note that for $m=2$ we obtain even the optimal constant 
(these are the classical inclusion relations for balls in Schatten classes).

\medskip

Let $\Hi=(\C^D)^{\otimes N}$ and $\Gamma(\Hi) \subset
\mathcal{B}(\Hi)$ be the convex hull of rank one product operators
$$\Gamma(\Hi):=\co\{|x_1\otimes\dots\otimes x_N\rangle\langle
y_1\otimes\dots\otimes y_N| \ :\ x_1,y_1,\ldots, x_N,y_N \in B_2^D\} .$$
The convex body $\Gamma(\Hi)$ is most naturally seen as the $N$th projective
tensor power of the set of (not necessarily Hermitian) operators on $\C^D$
whose trace class norm is
$\le 1$.  However, it can also be identified with
$(B_2^D)^{\hat{\otimes}2N}$
when we identify $\mathcal{B}(\mathcal{K})$ with $\overline{\mathcal{K}}
\otimes
\mathcal{K}$. It then follows from the preceding lemma that $\Gamma(\Hi)$
contains a Euclidean ball (a Hilbert-Schmidt ball in this context) of radius
$D^{-(2N-1)/2}$.

The next lemma will relate $\Gamma(\Hi)$ to $\Sigma(\Hi)$, allowing to
deduce that the latter body contains a suitably large Euclidean ball.
Let $d_N$ be the geometric distance
between the sets $\Delta$ and $\Sigma$ corresponding to $N$ qubits, i.e. the
smallest positive number such that $\Delta((\C^2)^{\otimes N}) \subset
d_N\Sigma((\C^2)^{{\otimes} N})$. We then have

\begin{lemme}
\label{proj} Let $\pi : \mathcal{B}(\Hi) \to \saop$ be the
projection onto Hermitian part, $\pi(A) := \frac{1}{2}(A+A^\dagger)$.
Then
$$ \pi(\Gamma(\Hi)) \; \subset \; d_N \, \Sigma(\Hi) .$$
\end{lemme}

{\bf Proof:} It is enough to show that extreme points of $\pi(\Gamma)$
are contained in $d_N \Sigma$. Any extreme point $A$ of
$\pi(\Gamma)$ can be written as
$$ A = \frac{1}{2} \left(|x_1\otimes\dots\otimes x_N\rangle\langle
y_1\otimes\dots\otimes y_N | +|y_1\otimes\dots\otimes
y_N\rangle\langle x_1\otimes\dots\otimes x_N |\right)$$ 
It may appear at the first sight that the above representation 
shows that $A$ is separable. However, while the two terms in the
parentheses are indeed product operators, they are not Hermitian and we can
only conclude that $A \in \Delta(\Hi)$ (as a Hermitian operator whose
trace class norm is $\le 1$).

 Let $\Hi_i$ be the 2-dimensional subspace of $\C^D$ spanned by
$x_i$ and $y_i$ (if the vectors are proportional, add any vector 
to get a 2-dimensional space) and let $\Hi' := \Hi_1 \otimes
\dots \otimes \Hi_N$. Then $A$ can be considered as an operator 
on $\Hi'$; more precisely, as an element of $\Delta(\Hi')$
(and, conversely, any operator acting on $\Hi'$ can be
canonically lifted to one acting on $\Hi$). 
Note that since we are in the asymptotics where
$N$ is small and $D$ is large, the dimension of the problem has been 
dramatically decreased.  Since $A$ belongs to
$\Delta(\Hi')$, it also belongs to $d_N \Sigma(\Hi')$, and thus
to $d_N \Sigma(\Hi)$.
\qed

\medskip

To effectively apply  Lemma \ref{proj},  upper bounds on $d_N$ are
needed. It has been observed in \cite{szarek}, Appendix H, that $d_N \le
6^{N/2}$. This estimate has been subsequently slightly improved in \cite{gb3} 
to $2/3 \times 6^{N/2}$, and even to $a_N \times 6^{N/2}$
with $\lim_{N\to\infty} a_N \approx 0.624$. [These bounds 
are in fact based on estimates for the {\em a priori }
larger quantities, the geometric distances between 
$\Sigma((\C^2)^{\otimes N})$
(or $\mathcal{S}((\C^2)^{\otimes N})-\Id/2^N$) and Hilbert-Schmidt balls,
which are essentially reciprocals of the Hilbert-Schmidt inradii of these
sets; and in this  latter context they are ``nearly optimal;"
see Theorem 3 and the comments following it.]   Thus
$\Sigma(\Hi) \supset 3/2 \times 6^{-N/2} \pi(\Gamma(\Hi))$; 
combining this inclusion with Lemma  \ref{inradius} or, more precisely,
with the observation following the definition of $\Gamma(\Hi)$, we conclude
that $\Sigma(\Hi)$ contains the Hilbert-Schmidt ball of radius 
$3/2 \times 6^{-N/2} D^{-(2N-1)/2} = 3/2 \times 6^{-N/2}/d^{1-1/2N}$.
This gives a lower bound for the volume radius
which, together with (\ref{voldd}), yields
$$
\left(\frac{\vol(\Sigma)}{\vol(\Delta)}\right)^{1/d^2}
\geq  \frac{3/4 \times 6^{-N/2}} {d^{1/2-1/2N}}
$$
To conclude the proof of the theorem  we pass to the bodies $\mathcal{S}$
and $\mathcal{D}$ using (\ref{r-s}).

\medskip

{\bf Remarks:} (1) The bipartite case ($N=2$) was studied in \cite{gb1},
which contains in particular the remarkable result that $\mathcal{D}$ and
$\mathcal{S}$ have then the same inradius, which equals
$1/\sqrt{d(d-1)}$. However, the inradii of the symmetrized bodies, which
are (in both cases) comparable with the volume radii, are
(asymptotically) very different: the inradius of $\Delta$ equals
$1/\sqrt{d}$, while the inradius of $\Sigma$ has just been shown to be of
order $1/d^{3/4}$.

(2) 
The upper bound on $d_N$ was further slightly improved in subsequent
versions of \cite{gb3} and, most recently, in \cite{ASW}, where it was shown
that  $\mathcal{S}$ contains a ball of radius $\sqrt{3} \times 6^{-N/2}$, 
which implies $d_N \le (3^{N-1}(2^N-1))^{1/2}$.

\section{Proof of Theorem 2: Large number of small subsystems}

This part deals with asymptotic estimations of volumes when $N$, the number
of subsystems, tends to infinity whereas the dimension $D$ of each subsystem
remains bounded. This case is probably the one with most physical
interest. When all subsystems are 2-dimensional (qubits), Theorem 2 has
been proved in \cite{szarek}.

For the proof of the present case it is more convenient to deal with
an affine image of $\Sigma$ which is {\em more balanced} than $\Sigma$
itself   or, equivalently, to consider a new scalar product which is better
adapted to the analysis of $\Sigma$  than the Hilbert-Schmidt product. It is
a known phenomenon that high-dimensional convex bodies often enjoy hidden
symmetries (or approximate symmetries) that are only revealed when one looks
at them through a suitable Euclidean structure. In our situation, in the
asymptotics when
$N$ is large and $D$ is bounded, our Theorem 1, obtained with the usual
Euclidean structure, misses the genuine volumic behavior of the bodies
$\mathcal{S}$ and
$\Sigma$. The appropriate scalar product will be here the one derived
from the L\"owner ellipsoid of $\Sigma$. To see why, just look back at
the equation (\ref{upper}). This bound is the tightest when $\mathcal{E}$ 
is the L\"owner ellipsoid of $\Sigma$. It turns out that it is possible
to describe completely this ellipsoid. First recall that
$\Sigma(\Hi) = \Delta (\C^D) \hat{\otimes} \dots \hat{\otimes}
\Delta (\C^D)$. By Lemma \ref{lemtens}, it is enough to calculate
the L\"owner ellipsoid of $\Delta(\C^D)$, and by tensoring we
get the L\"owner ellipsoid of $\Sigma(\Hi)$.

We will determine the L\"owner ellipsoid of $\Delta(\C^D)$ using
the following elementary general lemma.

\begin{lemme}
\label{lowsym} Let $u$ be a unit vector in $\R^n$ and
$H_0=\{u\}^\perp$ be the orthogonal hyperplane. Let $h>0$ be real
number and $H=H_0+hu$ be a translate of $H_0$. We consider a
convex body $\Omega= \co \{W \cup -W\}$, where $W$ is a convex
body in $H$. Then the following assertions are equivalent:
\begin{enumerate}
\item $\Omega$ is in L\"owner position (i.e. the unit ball $B_2^n$
is the L\"owner ellipsoid of $\Omega$). \item $h=1/\sqrt{n}$ and
$B_2^n \cap H$ is the L\"owner ellipsoid of $W$.
\end{enumerate}
\end{lemme}

{\bf Proof: } Assume that $\Omega$ is in L\"owner position, so
$\Omega \subset B_2^n$. By John's theorem (Proposition
\ref{john}), this means that there exists a John resolution of
identity $(x_i,c_i)$, with $x_i \in \Omega \cap S^{n-1}$ and $\sum
c_i |x_i\rangle\langle x_i|=\Id$. Since $x_i$ are extreme points
of $\Omega$, they must be in $W$ or $-W$, and since $\Omega$ is
symmetric we can assume they are all in $W$. Let $P$ be the
orthogonal projection onto $H_0$. It follows that  $\sum c_i
|Px_i\rangle\langle Px_i|=P$; note that $\|Px_i\|=\sqrt{1-h^2}$.
This means that $(Px_i,c'_i)$ is a John resolution of identity in
$H_0$, with $c'_i=(1-h^2)c_i$. By Proposition \ref{john},
$(1-h^2)^{-1/2}PW$ is in L\"owner position, or equivalently $B_2^n
\cap H$ is the L\"owner ellipsoid of $W$. Moreover, we have the
conditions $\sum c'_i=n-1$ and $\sum c_i=n$, which force
$h=1/\sqrt{n}$.

The converse follows by retracing the above argument in the opposite
direction. \qed

\medskip

\begin{lemme}
\label{calcul} The L\"owner ellipsoid of $\Delta(\C^D)$ is
determined by the following scalar product defined for $A, B \in
\mathcal{B}_{sa}(\C^D)$ by
$$ \<A,B\>_{\Lo(\Delta(\C^D))} = (1+\frac{1}{D}) \tr (AB) - \frac{1}{D}
\tr(A)\tr(B) $$
\end{lemme}

\medskip

{\bf Proof:} First remark that, because of its uniqueness, the
L\"owner ellipsoid of any body $K$ must inherit all the symmetries
of $K$. Since $\mathcal{D}(\C^D)$ is invariant under unitary conjugations
(i.e., the maps $\pi_U(X) = U X U^\dagger$ with $U \in U(D)$), the
ellipsoid
$\Lo(\mathcal{D}(\C^D))$ must be similarly invariant, and thus it is
the  circumscribed Euclidean ball. This is actually
somewhat delicate; the relevant fact is that the action  of the group $\{
\pi_U \, :
\,  U \in U(D) \}$  on the hyperplane $\mathcal{T}_1$ containing
$\mathcal{D}(\C^D)$ is irreducible (in particular, $\Id/D$ is the only fixed
point and so it must be the center). In the language of convex bodies (see
\cite{nicolebook}) we say that
$\mathcal{D}(\C^D)$ has enough symmetries.
On the other hand, the fact that this group acts transitively on the set
of extreme points of $\mathcal{D}(\C^D)$ (which are the pure states)
implies {\em by itself} only that all these extreme points are contact
points, i.e., belong to the boundary of the L\"owner ellipsoid.

By Lemma \ref{lowsym}, we know that the trace of the L\"owner
ellipsoid of $\Delta(\C^D)$ on the hyperplane $\mathcal{T}_1$ is a
Euclidean ball (centered at $\Id/D$), so we have for some real numbers
$\alpha$ and
$\beta$
$$\<A,B\>_{\Lo(\Delta(\C^D))} = \alpha \tr (AB) + \beta \tr(A)\tr(B)$$
To determine $\alpha$ and $\beta$, observe that
$\<\rho,\rho\>_{\Lo(\Delta(\C^D))}=1$ for any pure state (hence
contact point) $\rho$, and that (from the condition on $h$ in Lemma
\ref{lowsym}; note that $n=D^2$),
$\<\Id/D,\Id/D\>_{\Lo(\Delta(\C^D))}=1/D^2$. This gives $\alpha =
1+1/D$ and $\beta=-1/D$. \qed

\medskip

It is possible to use the L\"owner ellipsoid of $\Sigma(\Hi)$ as
the ellipsoid $\mathcal{E}$ in formula (\ref{upper}). Choosing
$\delta = 1/\sqrt{N \ln 2N}$ as earlier, we obtain the upper estimate
in the formula
\begin{equation}
\label{heart} \frac 1d\leq \left( \frac{\vol(\Sigma)}{\vol(\Lo
(\Sigma))} \right)^{1/d^2} \leq \;
\frac{C \sqrt{DN\ln N}}{d}
\end{equation}
The lower estimate follows from the classical fact (\cite{ball})
that for a centrally symmetric convex body $K$ in $\R^n$, we have
the inclusion $\Lo (K) \subset \sqrt{n} \, K$. Improvements on
constants can be derived from a result of Barthe (\cite{barthe}) which
asserts that for an $n$-dimensional symmetric body $K$, the ratio
$(\vol(K)/\vol(\Lo(K))^{1/n}$ is minimal when $K$ is the unit ball
of $\ell_1^n$ (and thus is {\em at least }
$(4/\pi)^{1/2}(\Gamma(n/2+1)/\Gamma(n+1))^{1/n}) = 
\sqrt{\frac{2e}{\pi}}(1-O(\frac{1}{n})) \frac{1}{\sqrt{n}}$ ; we recall that
here $n=d^2$).

\smallskip
It remains to calculate the volume of the L\"owner ellipsoid of
$\Sigma$. It follows from Lemma \ref{lemtens} that, if $\Phi$
denotes a linear map which sends $B_2^{D^2}$ onto $\Lo
(\Delta(\C^D))$, then $\Psi:=\Phi^{\otimes N}$ maps $B_{HS}$ onto
$\Lo(\Sigma(\Hi))$. By Lemma \ref{calcul}, we can define $\Phi$ by
$\Phi(A)=(1+1/D)^{-1/2}A$ if $A$ has trace $0$ and
$\Phi(\Id)=\sqrt{D} \, \Id$. This gives
$\det(\Phi)=\sqrt{D(1+1/D)^{1-D^2}}$ and finally 
\begin{equation}
\label{final}  \frac{\vol(\Lo (\Sigma))}{\vol(B_{HS})}
 = \det (\Psi) = \det(\Phi)^{N D^{2N-2}} = d^{-\alpha_Dd^2}
\end{equation}
To complete the proof it remains to combine the volume
estimates (\ref{voldd}),(\ref{heart}),(\ref{final}) and
(\ref{r-s}).

\section{Proof of Theorem 3: Tight upper bounds on the inradii}  

The idea is to show that there is a projection of the set 
$\Sigma = \Sigma((\C^2)^{\otimes N})$ which is demonstrably small. 
Specifically, let 
$P : \mathcal{B}_{sa}(\C^2) \rightarrow \mathcal{B}_{sa}(\C^2)$ 
be the orthogonal projection onto the 3-dimensional subspace of trace zero 
matrices and let $\Pi := P^{\otimes N}$. Since projective tensor products 
commute with linear maps (in the sense that 
$(T_1 \otimes \ldots \otimes T_s)(K_1
\hat{\otimes} \ldots \hat{\otimes} K_s) = (T_1K_1)
\hat{\otimes} \ldots \hat{\otimes}(T_sK_s)$), we have 
$$
\Pi\left(\Sigma((\C^2)^{\otimes N})\right) =  
P^{\otimes N} \left(\Delta(\C^2)^{\hat{\otimes}N}\right) 
= \left( P (\Delta(\C^2))\right)^{\hat{\otimes}N} .
$$
Since $\Delta(\C^2)$ is a cylinder whose axis coincides with the kernel of $P$, 
the image of $\Delta(\C^2)$ under $P$ is a 3-dimensional Euclidean ball 
of radius $1/\sqrt{2}$  (more precisely, $\mathcal{D}(\C^2) - \Id/2$).  
Accordingly, the image of the corresponding set $\Sigma$ under $\Pi$ 
is the $n$th projective tensor power of such a ball. The relevant parameters 
of such sets (mean widths, volume radii) are majorized by the following
statement,  which is a complement of Lemma 4.

\begin{lemme}
\label{vrad} 
There is an absolute constant $C_1$ such that if $D, m$ are positive
integers $(m\geq 2)$ and 
$K := (B_2^D)^{\hat{\otimes} m}$ is the projective tensor product of $m$ 
copies of the  $D$-dimensional Euclidean ball (real or complex), then
$$ \vrad(K) \leq w(K) < \frac{C_1 \sqrt{m \ln{m}}}{D^{(m-1)/2}} $$
\end{lemme}
{\bf Proof:}
The first inequality in Lemma \ref{vrad} -- comparing the volume radius 
of a convex body to its mean width -- is just the Urysohn
inequality. The proof of the second follows the same lines as
 -- but is simpler than
-- the approach that yielded formula  (\ref{upper}). 
We present the details only for the
case that is relevant to Theorem 3, namely $D=3$ in the real setting. The
argument is very similar to the one used in \cite{szarek} to obtain an upper
bound for the volume of $\Sigma((\C^2)^{\otimes m})$, where the reader is
referred for additional background and justifications.

We start by noting that, for any $\delta \in (0, \sqrt{2})$, the sphere $S^2$
contains a $\delta$-net $\mathcal{F}$ whose cardinality is $\leq 16/\delta^2$. On
the other hand, an elementary geometric argument shows that the convex hull of
any $\delta$-net contains a (Euclidean) ball of radius $1-\delta^2/2$.
Accordingly, the convex hull of $\mathcal{F}^{\otimes m}$, the $m$th tensor
power of $\mathcal{F}$, contains  $(1-\delta^2/2)^m(B_2^3)^{\hat{\otimes} m}$.
Since 
$\# (\mathcal{F}^{\otimes m}) \leq (\# \mathcal{F})^m \leq (4/\delta)^{2m}$, 
it follows from (\ref{volpol}) that 
$$
w((1-\delta^2/2)^m(B_2^3)^{\hat{\otimes} m}) \leq w( \co
(\mathcal{F}^{\otimes m})) < 
\gamma_{3^m}^{-1} \sqrt{2 \ln{((4/\delta)^{2m})}}
$$
and, consequently,
$$
\vrad((B_2^3)^{\hat{\otimes} m})\leq w((B_2^3)^{\hat{\otimes} m}) 
<2\gamma_{3^m}^{-1}
(1-\delta^2/2)^{-m}\sqrt{ m\ln{(4/\delta)}} .
$$
We now choose $\delta = 1/\sqrt{m \ln{2m}}$; then, for large $m$,  
$(1-\delta^2/2)^{-m} \sim 1$, $\ln{(4/\delta)} \sim \frac 12 \ln{m}$ and 
$\gamma_{3^m} \sim 3^{m/2}$. This yields the asserted upper bound on the 
mean width with $C_1 \sim \sqrt{2/3}$ for large $m$. A more careful 
calculation using Lemma \ref{tenswidth} for small $m$ shows that 
$C_1 = 1.673$ works for all  $m$. [The discussion of constants 
above and in the remainder of this section is specific to $D=2$, but
can be analogously carried over in the general case; for example, it is
easy to show   that $C_1 \sim 1$ works for large $m$  and all $D$.]
As in \cite{szarek}, further improvements are possible by working with 
explicit nets and/or by noting that $\mathcal{F}$ can be chosen to be 
symmetric, but we will not pursue them since it is likely that,  for the
problem at hand,  sharper estimates can be obtained by analytic methods.
\qed

\medskip

We now return to the proof of Theorem 3. From prior considerations,
the image of $\Sigma$ under the projection $\Pi$ is congruent to 
$2^{-N/2} (B_2^3)^{\hat{\otimes} N}$ and so, by Lemma \ref{vrad}, 
$$
w(\Pi(\Sigma)) = 2^{-N/2} w((B_2^3)^{\hat{\otimes} N} 
< 3^{1/2}C_1 \sqrt{N \ln{N}} \; 6^{-N/2} .
$$
Since the inradius is trivially smaller than the mean width, and 
since the inradius of the set $\mathcal{S}$ of separable states  
does not exceed that of its symmetrization $\Sigma$, 
Theorem 3 follows (with $C_0= 3^{1/2}C_1 < 3$). It is also routine to obtain 
an (identical or nearly identical) upper bound on the mean width of
$\mathcal{S}$; however, some care is  needed because the dimensions of
$\mathcal{S}$ and $\Sigma$ are not the same.

\medskip
Since our argument is of a global nature, we are not able to produce an 
explicit state in the ball of radius $r:=C_0\sqrt{N\ln{N}} \; 6^{-N/2}$ 
and centered at $\Id/d$ which is not separable. However, standard 
tools of asymptotic geometric analysis (cf. \cite{st} or \cite{pisier})
imply that even stronger properties 
are satisfied ``generically." Denote by $\Hi_0$ the 
$d_0:=3^N$-dimensional range of the projection $\Pi$. Then, for most of 
the unit vectors $u$ on the unit sphere of $\Hi_0$, the state $\sigma 
= \Id/2^N + 2ru$ is not separable, and the same holds for any state 
$\rho$ such that $\Pi(\rho)= \Pi(\sigma)$ (here ``most of" means
that the  exceptional set is of normalized measure $< 2^{-d_0/2})$. Moreover, 
a slightly weaker but comparable property holds {\em simultaneously } 
for all  $u$ in a  ``generic" (in the sense of the invariant measure on the 
corresponding Grassmannian) $\lceil d_0/2 \rceil$-dimensional subspace 
of $\Hi_0$.

As mentioned in  the introduction, the above argument can be
generalized to other values of $D$ and leads to an upper bound for the 
inradii of the sets $\Sigma$ and $\mathcal{S}$ corresponding to 
$(\C^D)^{ \otimes N}$ which is, up to a logarithmic factor,
$(D(D+1))^{-N/2}$.  While this upper bound differs, for $D>2$, from the lower
bound  for the inradius of $\mathcal{S}$ that was found in \cite{gb3}, it
gives the correct order for the inradius of the set $\Sigma$. Indeed,
$\Sigma$ contains
$\Lo (\Sigma)/d$, which itself contains $(D(D+1))^{-N/2} B_{HS}$ 
(this argument parallels the one for qubits given in \cite{szarek}, 
Appendix H, where we refer the reader for details).

\section{Proof of Theorem 4: Asymptotic roughness of PPT criterion}

We will use the following proposition which is taken from
\cite{mp}.

\begin{proposition}
Let $K$ and $L$ be convex bodies in $\R^n$ with the same centroid.
Then
$$ \vol(K) \vol(L) \leq \vol(K-L)\vol(K \cap L),$$
where $K-L := \{x-y\, : \, x \in K, y \in L \}$ is the Minkowski 
difference.
\end{proposition}

\smallskip

We apply the proposition with $K=\mathcal{D}$ and
$L=T\mathcal{D}$. This gives

\begin{equation}
\frac{\vol(\mathcal{D})}{\vol(\mathrm{PPT})} \leq
\frac{\vol(\mathcal{D}-T\mathcal{D})}{\vol(\mathcal{D})}
\label{ineqmp}
\end{equation}

Note that we actually used the proposition in the space 
$\mathcal{T}_1 \sim \R^n$, seen as
a vector space with $\Id/d$ as origin; the point $\Id/d$ is also the
centroid of $\mathcal{D}$ (and $T\mathcal{D}$), it is even the 
the only fixed point under the group of symmetries of
$\mathcal{D}$ (cf. the proof of Lemma \ref{calcul}). We rewrite
(\ref{ineqmp}) using volume radii

$$ \left(\frac{\vol(\mathcal{D})}{\vol(\mathrm{PPT})}\right)^{1/n}
\leq \frac{\vrad(\mathcal{D}-T\mathcal{D})}{\vrad(\mathcal{D})} $$

Recall that we majorized the volume radius by the mean width using 
the Urysohn inequality
(here mean width is also taken in $\mathcal{T}_1$). This is very
convenient since, as one can check immediately from the definition, the
mean width is additive with respect to the Minkowski operations:
$w(\mathcal{D}-T\mathcal{D})=w(\mathcal{D})+w(-T\mathcal{D})$. Since
$-T$ is an isometry, it preserves the uniform measure on the sphere
$S^{n-1}$ and thus $w(-T\mathcal{D})=w(\mathcal{D})$. This gives

$$ \left(\frac{\vol(\mathcal{D})}{\vol(\mathrm{PPT})}\right)^{1/n}
\leq \frac{2w(\mathcal{D})}{\vrad(\mathcal{D})} $$

We now appeal to (\ref{vold}) to majorize the right hand side by 8, 
which proves Theorem 4 with $c_0 =1/8$.
Note also that one can use the explicit formula (\ref{explicit}) for the
volume of $\mathcal{D}$; this shows that asymptotically (the lower estimate
for) $c_0$ tends to
$e^{-1/4}/4 \approx 0.195$. Moreover, numerical evidence (which presumably can
be confirmed by rigorous calculation using appropriate 
Stirling-type formulae, or inequalities similar to those of
\cite{luke}) suggests that $\vrad(\mathcal{D}) \geq e^{-1/4}/\sqrt{d}$
for {\em all } dimensions
$d$, which would imply that the asymptotic value is in fact a bound.

Finally, observe that the above proof works actually for any
(Hilbert-Schmidt) isometry in $\mathcal{T}_1$ which fixes $\Id/d$,
and not only for the partial transpose $T$.

\medskip \noindent {\em Acknowledgments:} The contribution 
of the first named author
is a part of his Ph.~D.~thesis written at the University of
Paris VI under the supervision of the second named author,
and has been partially supported by the European Network PHD, MCRN-511953.
The second named author has been partially supported by the National Science 
Foundation (U.S.A.).

\vskip 16pt

{\small
\noindent {\'{E}quipe d'Analyse Fonctionnelle, B.C. 186,
Universit\'{e} Paris VI, 4 Place Jussieu,  F-75252  Paris, France\\
E-mail: \verb!aubrun@math.jussieu.fr!

\vskip 8pt

\noindent Department of Mathematics,
Case Western Reserve University,
Cleveland, OH 44106-7058, U.S.A.

\noindent {\em and }

\noindent 
\'{E}quipe d'Analyse Fonctionnelle, B.C. 186,
Universit\'{e} Paris VI, 4 Place Jussieu,  F-75252  Paris, France\\
E-mail: \verb!szarek@cwru.edu!
}}

\end{document}